\documentstyle[11pt,paspconf,psfig]{article}

\begin{document}

\title{Large Scale Structures in the Las Campanas Redshift Survey and in 
Simulations}

\author{V. M\"uller}
\affil{Astrophysikalisches Institut Potsdam, An der Sternwarte 16, 
       D--14482 Potsdam, Germany}
\author{A.G. Doroshkevich}
\affil{Theoretical Astrophysics Center, Juliane Maries Vej 30, 
       DK--2100 Copenhagen \O, Denmark}
\author{J. Retzlaff}
\affil{Max-Planck-Institut f. extraterrestrische Physik,
       Giessenbachstrasse, D--85740 Garching}
\author{V. Turchaninov}
\affil{Keldysh Institute of Applied Mathematics, Russian Academy of 
       Sciences, 125047 Moscow, Russia}

\begin{abstract}

The large supercluster structures obvious in recent galaxy redshift surveys are
quantified using an one-dimensional cluster analysis (core sampling) and a
three-dimensional cluster analysis based on the minimal spanning tree.  The
comparison with the LCRS reveals promising stable results.  At a mean
overdensity of about ten, the supercluster systems form huge wall-like
structures comprising about 40\% of all galaxies.  The overdense clusters have a
low mean transverse velocity dispersion of about 400 km/s, i.e.  they look quite
narrow in redshift space.  We performed N-body simulations with large box sizes
for six cosmological scenarios.  The quantitative analysis shows that the
observed structures can be understood best in low density models with $\Omega_m
\le 0.5$ with or without a cosmological constant.

\end{abstract}


\keywords{Cosmology:observations-Cosmology:large-scale structure 
of the Universe }

\section{Introduction}

Very large structures in the universe as the great attractor (Dressler et al.
1988) or the great wall (Ramella, Geller, \& Huchra 1992) were firstly
considered as rare peculiarities in the large scale galaxy distribution.  But
recently published rich galaxy redshift catalogues as the Durham/UKST Galaxy
Redshift Survey (Ratcliffe et al.  1996) and the Las Campanas Redshift Survey
(LCRS) (Schectman et al.  1996) have established the existence of similar large
structures as a typical characteristic of the galaxy distribution.  They
correspond to the superclusters first identified in the distribution of clusters
of galaxies (Abell, 1958; Oort, 1963), and we shall call them in the following
simply superclusters.  These structures exhibit a clear filamentary network, and
the filaments seem to occupy very large wall-like regions (Doroshkevich et al.
1996).  The high density superclusters incorporate $\sim$40\% of all galaxies,
and they have a typical diameter of $\sim$30$h^{-1}$Mpc (Hubble parameter $h$ in
units of 100 km s$^{-1}$ Mpc$^{-1}$).  They surround huge underdense regions of
diameter (50 -- 70)$h^{-1}$Mpc.

The appearance of such a supercluster distribution is a typical outcome of the
nonlinear gravitational instability as described by the Zel'dovich theory (e.g.
Shandarin \& Zel'dovich, 1989).  The filamentary galaxy distribution has been
characterized as forming a cosmic web with a sponge like structure (Bond et al.
1996).  A quantitative analysis shows that the characteristics of very large
structures and the evolution history are quite sensitive to the specific
cosmological model.  In particular, the wall like structures are formed during a
short evolution stage, and later they disintegrate into a system of high density
clumps (Doroshkevich et al.  1997).  The large scale matter distribution can be
traced back to the initial peculiar gravitational potential (Madsen et al.
1998).  Using statistical characteristics of the potential in different
cosmological scenarios, the formation and evolution of the supercluster
structures can be described quantitatively (Demia\'nski \& Doroshkevich 1998).
These theoretical predictions and our numerical experiments can be used to
constrain cosmological scenarios which give rise to the various peculiar
potential fields.

\begin{figure}
\centerline{\vbox{
\psfig{figure=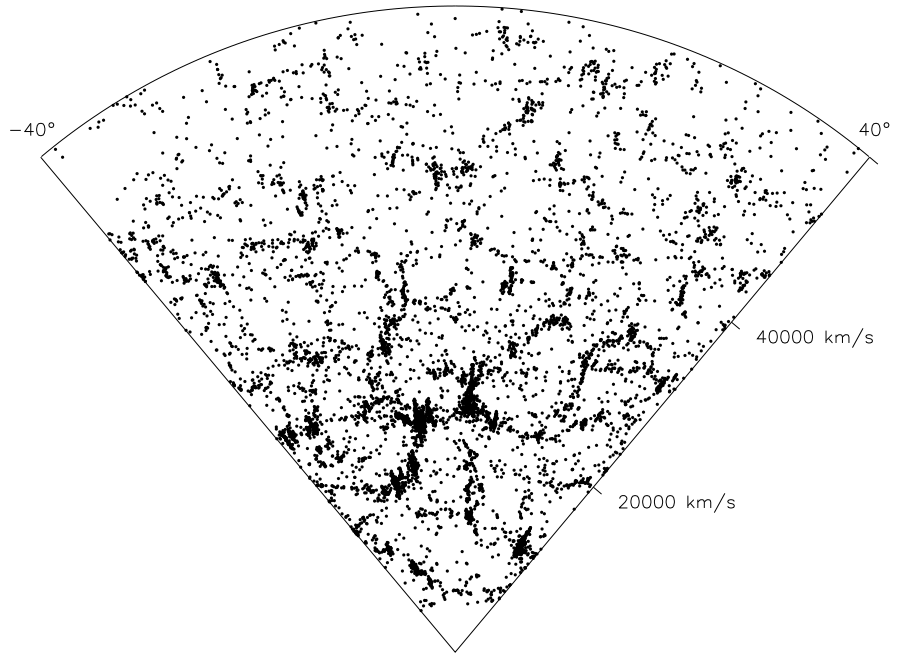,width=1.0\hsize}
}}
\vspace{-1.2cm}
\caption[]{Wedge diagram of $80^{\circ}$ width and $1.5^{\circ}$ depth of a 
mock catalog for the magnitude limited 
LCRS (7500 galaxies with $15 < m < 18$.)}
\end{figure}

We concentrate on a quantitative characterization of the geometrical properties
of the supercluster structures.  We take up the attempts of Babul \& Starkman
(1992) and Luo \& Vishniac (1995) to quantify the linearity and planearity of
the galaxy clustering with moments of the tensor of inertia.  These statistics
alone do not strongly discriminate between different cosmological models, since
they characterize the general character of the anisotropic gravitational
collapse.  We perform a cluster analysis separately in high-density and
low-density regions of the universe (the demarkation line lies about at an
overdensity 10, see below), and we provide characteristics of the clusters and
superclusters in dependence of the linking length used to define it.  Our basic
results lead to a remarkable discrimination between rich structure elements and
low density regions in the universe.  Recently, specific filament finders are
investigated (Dav\'e et al.  1997) which are able to follow the cosmic web.
This statistic is strongly restricted to the small scale galaxy distribution.
It discriminate marginally between different scenarios of structure formation,
but no specific model is clearly favoured in the analysis of the CfA redshift
survey.  Here we employ the core sampling approach and the minimal spanning tree
to define and to characterize the wall-like and the filamentary structures and
to measure the parameters of their sizes, abundances and distributions.

First we provide a description of the simulations for modeling the supercluster
distribution.  Next we describe the cluster analysis and minimal spanning tree
techniques used to analyse it.  Then we employ the core sampling approach to
characterize the supercluster structures in comparison to the LCRS.  Finally we
compare quantitative characteristics of the supercluster structures in
dependence on theoretical predictions, and we discuss the results.  More details
can be found in Doroshkevich et al.  (1998).

\section{Simulating the large scale structure}

We study the large scale matter distribution in N-body simulations using some of
the promising cosmological scenarios presently under discussion.  We use six
cosmological models:  A COBE normalized CDM model with density parameter
$\Omega_m=1$ (SCDM) is taken as a comparison model despite of its well known
difficulties.  Alternatives based on CDM and a flat geometry of the universe
include modifications of the primordial power spectrum by introducing a tilt
$\propto k^{0.9}$, or a break in the power spectrum at a certain scale, TCDM and
BCDM, respectively.  Both are inflation motivated (Gottl\"ober et al.  1991,
1994).  Further we use open CDM models with $\Omega_m=0.5$ and $\Omega_m=0.35$,
OCDM1 and OCDM2.  Furthermore we take a model with $\Omega_0=0.35$ and a
cosmological constant to provide spatial flatness, $\Lambda$CDM.  All models are
COBE normalized (Gorski et al.  1994, Stompor 1995), the amplitude is
characterized by the mass variance $\sigma_8$ on a scale of 8$h^{-1}$Mpc as
given in Table 1.

\begin{table}
\caption{Parameters of simulations.} \label{tbl1}
\begin{center}\scriptsize
\begin{tabular}{cccccccc} 
model&$\Omega_m$&$h$&$m_{p}$ &$\sigma_8$&$\sigma_{vel}$&$\tau_0$&$l_0$ \cr
 & & &$10^{11}{\rm{M_\odot}}$&        & km/s       & &$h^{-1}$Mpc\cr

SCDM         & 1~~~ & 0.5~ & 3.2  & 1.37 & 1375 & 13.2 & 0.6~ \cr
TCDM~~       & 1~~~ & 0.5~ & 3.2  & 1.25 & 1293 & 13.2 & 0.57 \cr
BCDM~~       & 1~~~ & 0.5~ & 3.2  & 0.60 & ~714 & 13.2 & 0.31 \cr
OCDM1~~      & 0.5~ & 0.6~ & 1.6  & 0.74 & ~550 & 22.~ & 0.14 \cr
OCDM2~~      & 0.35 & 0.65 & 1.1  & 0.57 & ~372 & 29.~ & 0.14 \cr
$\Lambda$CDM & 0.35 & 0.7~ & 1.1  & 1.12 & ~913 & 26.9 & 0.2~ \cr
\end{tabular}
\end{center}
\end{table}

The simulations were run in boxes of very large comoving side length
$L_{box}=500h^{-1}$Mpc to include with high statistics the range of wave numbers
$k^{-1}\approx$ (10 -- 30) $h^{-1}$Mpc responsible for the supercluster
formation.  We use a PM code, described in more detail in Kates et al.  (1995).
We use $N_{p} = 300^3$ particles in $N_{cell}=600^3$ grid cells, which provides
a force resolution $\sim 0.9h^{-1}$Mpc, the mass resolution is given by $m_{p}$
in the Table 1.  For later reference, we show also the velocity dispersion of
the matter particles $\sigma_{vel}$ in the different simulations, which are
scaled to the present time according to the linear perturbation theory.  A
length scale $l_0$ and a dimensionless constant $\tau_0$ characterize the
typical scale of the supercluster structures, and their evolution stage as
discussed below:  $$ l_0^{-2} = \int_ {k_{min}}^{k_{max}}k^1~T(k)~dk, \quad
\tau_0 = {\sigma_{pvel}\over \sqrt{3} l_0 H_0}.  $$ Here $k$ is the comoving
wave number, the factor $k^1$ stems from the input primordial
Harrison-Zel'dovich spectrum, $T(k)$ denotes the transfer function of
perturbations in the different scenarios, and $k_{min}=2\pi/L_{box}$, and
$k_{max}=k_{min}N_{cell}^{1/3}$ gives the range of the spectrum realized in the
simulations.

In Retzlaff et al.  (1998), we employed these simulations to discuss the power
spectrum of the clustering of Abell clusters.  There, model clusters were
identified with maxima in the smoothed density field.  Now we use the dark
matter particles to characterize the geometry of the superclustering.  For
comparison with the observed galaxy catalogues, we must allow for differences in
the distribution of galaxies and dark matter.  Here we start with the simple
hypothesis that on large scales, the galaxy distribution follows the gross
distribution of the dark matter.  As shown by the mass variance of the different
models, we must allow for a bias or an anti-bias in the small scale galaxy
clustering to get the observed galaxy variance $\sigma_8(\mbox{gal}) = 1$, and
to reproduce the observed galaxy-galaxy correlation function (Tucker et al.
1997).  However, we show results only for the full dark matter distribution.
They are independent of any additional assumptions on prescriptions for the
`galaxy identification'.  Effectively, we employ mass cuts in the cluster
definitions which are equivalent to a biasing.  For producing `galaxy
catalogues' we also employed simple local bias prescriptions depending on the
environmental density.  Our conclusions are basically unchanged by such a
biasing.  The reason is that we study the geometrical properties of the matter
distributions on very large scales which are independent on differences of the
clustering of dark matter and galaxies in high density clumps.

To illustrate the simulations, we show in Fig.  1 a cone diagram of a mock
`galaxy sample' for the LCRS which is based on $\Lambda$CDM.  We took the
Schechter function of the LCRS-galaxies and assigned randomly luminosities to
the dark matter mass particles according to the Schechter function
characterizing the LCRS (Lin et al.  1996).  Obviously, huge overdense regions
are seen which match properties of the real observations.

\section{Geometric characterization of large scale structures}

The cluster analysis is widely used in cosmology both to define virialised
objects, and for large linking lengths, to describe characteristics of the large
scale matter distribution up to percolation, cf.  e.g.  Sahni \& Coles (1995).
It is closely related to the characterization of the galaxy distribution by the
minimal spanning tree (MST, cf.  Barrow et al.  1985).  The MST represents an
unique geometric construction in the galaxy distribution which is based solely
on the galaxy positions.  The simplest characterization of the MST is the number
distribution of the edge length, $W_{MST}(l)$.  First, Barrow et al.  (1985)
constructed it for the CfA-catalogue.  At small separations, this distribution
characterizes the non-linear galaxy clustering, and on large scales, it provides
a measure of the topology of the geometrical support of the galaxy distribution.
For random particles situated on two-dimensional (walls) or one-dimensional
structures (filaments), we get logarithmic increments in the number of edge
length of two and one, respectively.  Parametrising the edge length distribution
by 
$$W_{MST}(x)=-W_0~{dF_f\over dx}~e^{-F_f(x)},~x=l/<l>,$$ 
where $l$ denotes the edge lengths of the MST and $<l>$ its mean.  $W_0$
provides the normalization and the function $F_f(x)$ is parametrised as
$$F_f(x)=(\beta_1x^{p_1}+\beta_2 x^{p_2})^{p_3}.$$ 
As mentioned, we are mainly interested in the effective power index 
$ p(x) = {{d \log F_f} / {d \log x}},$ characterizing the geometry of the galaxy
distribution.

Splitting off the MST at different linking lengths, we get standard
friends-of-friends clusters with various linking lengths $r_{link}$, the number
of clusters $N^{t}_{cl}(r_{link})$ (the exponent $t$ is used to indicate on
the total number of clusters, we discuss separately also clusters with different
multiplicity, singlets $N^{s}_{cl}$, doublets etc.):
$$N^{t}_{cl}(r_{link}) = \int_{r_{link}}^\infty W_{MST}(l)dl. $$

\begin{table}
\caption{Power law indices of cluster numbers and MST edge lengths 
in the $\Lambda$CDM model, TOT, RSE, and LDR denote total sample, rich structure 
elements, and low-dense regions, respectively; com in comoving and red in 
redshift space.} \label{tbl2}
\begin{center}\scriptsize
\begin{tabular}{ccccccc} 
 sample&$f_{p}$& $<n_{p}>$&$p^{s}$&$p^{t}$&$<l_{MST}>$&$p^{MST}$\cr
       &        &$h^3Mpc^{-3}$&       &       &$h^{-1}$Mpc&         \cr
TOT-com &1.0~ & 0.113 & $0.71\pm 0.02$&$0.59\pm 0.03$& 0.72&$0.60 \pm  0.02$\cr
TOT-red &1.0~ & 0.113 & $0.76\pm 0.02$&$0.85\pm 0.04$& 0.78&$0.91 \pm  0.03$\cr
RSE-red &0.45 & 1.86~ & $1.82\pm 0.02$&$1.83\pm 0.03$& 0.42&$1.60 \pm  0.04$\cr
LDR-red &0.55 & 0.065 & $1.01\pm 0.02$&$1.11\pm 0.02$& 1.00&$1.11 \pm  0.03$\cr
\end{tabular}
\end{center}
\end{table}

In Table 2 we provide the power law exponents of the edge length distributions
of the MST ($p^{MST}$) and of the cluster number versus linking length ($p^s$
for singlet clusters, $p^{t}$ for complete clusters) in simulated dark matter
distributions of the $\Lambda$CDM model as a typical example.  We show the
results in the physical space and, to allow a comparison with the data from
observed redshift catalogues, in redshift space, were the galaxy positions are
shifted along one axis according to their peculiar velocities.  In addition we
use a linking length $l_{link}$ of 1 $h^{-1}$Mpc and a threshold multiplicity
$N^{thr} = 200$ for discriminating rich structure elements (RSE) with a minimum
overdensity.  Effectively this determines those particles in the simulations
which mark `superclusters' in the galaxy distribution.  The remaining particles
are separately analysed, they form low-density regions (LDR) which are
characterized by field galaxies.  The same discrimination was made by
Doroshkevich et al.  (1996) in the Las Campanas Redshift Survey.  Similar as in
the galaxy data, we identify a percentage $f \approx 0.40$ in rich structure
elements in some cosmological models.  In the simulations we find power law
exponents $p^s$, $p^t$, and $p^{MST}$ of about unity if analysed in physical
space, i.e.  the matter distribution is basically filamentary.  In redshift
space, the power laws for rich structure elements have exponents $p^{t} \approx
1.6$ and $p^{MST} \approx 1.8$, while again they are near unity in low-density
regions.  This difference is the main justification for the discrimination
between different populations in the galaxy distribution.  In redshift space the
influence of the velocity dispersion erases the filamentary character of the
matter distribution in RSE.  Therefore we find an apparent particle distribution
in huge wall-like structures.

In the LCRS a power index $p_t \approx p_{MST}\approx 1.7$ has been found for
RSE, and $p_t \approx p_{MST}\approx 1$ for LDR and for the total sample.  This
is comparable with our results in redshift space.  A complex inner structure of
RSE is also visually seen in the LCRS which resamples the galaxy distribution in
the great wall of the CfA catalog (Ramella, Geller \& Huchra 1992).

\begin{figure}
\centerline{\vbox{
\psfig{figure=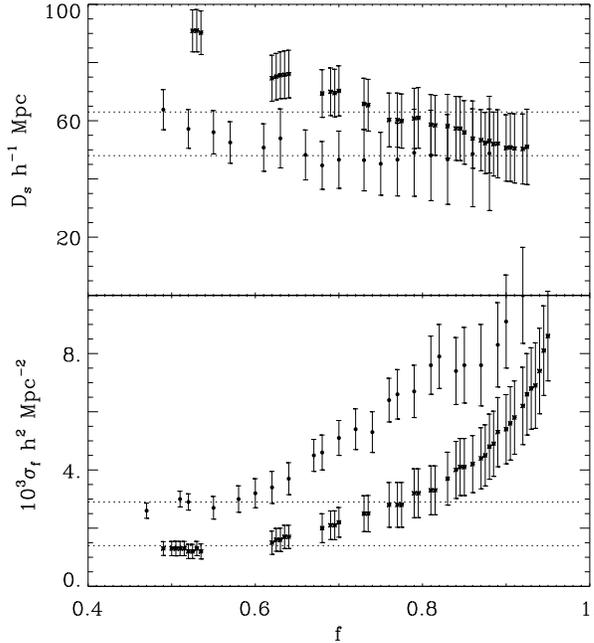,width=0.58\hsize}
}}
\vspace{0.5cm}
\caption[]{Mean separation of the RSE, $D_s$, and the
surface density of filamentary component, $\sigma_f$, vs.
the matter fraction concentrated within the structures, $f$,
in redshift space for the OCDM1 (dots) and
$\Lambda$CDM (stars) models.}
\end{figure}

\section{Mean separation of filamentary and sheet-like components}

The core-sampling method of Buryak et al.  (1994) allows us to discriminate the
filamentary and sheet-like structure elements and to find quantitative
characteristics of these structures, namely, the surface density of filaments,
$\sigma_f$, that is the mean number of filaments crossing a randomly oriented
unit area (in $h^{-2}$ Mpc$^2$), and the linear density of sheets, $\sigma_s$,
that is the number of sheets crossing the unit length (in $h^{-1}$ Mpc) of a
random straight line.  These parameters are equivalent to the mean separation
between sheet-like structure elements, $D_s$, and filaments, $D_f$:
$$D_s = 1/\sigma_s, \qquad D_f =\sigma_f^{-1/2}, $$
i.e.  these lengths represent the mean free path between sheet-like and
filamentary structure elements.

For our analysis, 196 cylinders with a radius of 1.7$h^{-1}$Mpc were prepared.
The mean number of points within the cylindrical cores amounts $\sim$(400 --
600).  The analysis was performed for 16 values of the cylinder radius,
$1.7h^{-1}$Mpc $\geq r_{cyl}\geq 0.7h^{-1}$Mpc.  The mean separation of
sheet-like elements, $D_s$, and the surface density of the filamentary component
$\sigma_f$, are plotted in Fig.  2 as function of the fraction of matter which
is randomly selected.  We provide results only for the OCDM1 and $\Lambda$CDM
models which are the only reasonable models as the next section demonstrates.
The range of values found in the LCRS (Doroshkevich et al.  1996) are marked by
dotted lines.  There is a clear signal from a wall-like supercluster population
with a characteristic separation of $D_S \approx (40 - 60)h^{-1}$ Mpc.  But in
contrast with results from the observations, $D_s$ increases slowly for small $f
\leq$ 0.6.  This effect is probably caused by differences in the covering of the
cores by the walls in the simulations, cf.  also Ramella et al.  (1992) and
Buryak et al.  (1994).  On the other hand, the nearly constant $D_s(f)$ and a
quick drop of $\sigma_f$ with $f$ proofs that a significant matter fraction
($\sim 0.4 - 0.5$) is associated with the high dense sheet-like component,
consistent with the observations.

\section{Quantitative description of the overdensity regions}

The discrimination of rich structure elements (RSE, or `superclusters') and
low-density regions (LDR populated by field galaxies) delivers no sharp
criterion, we varied the linking length and the mass threshold for identifying
structures.  The results are shown in Fig.  3 and 4 for the various cosmological
models.  As discussed below, the discrimination is much more dependent on the
type of cosmological model than on the exact boundary and criteria used.
Therefore, the analysis delievers a test of cosmological models which is very
robust.

\begin{figure}
\centerline{\vbox{
\psfig{figure=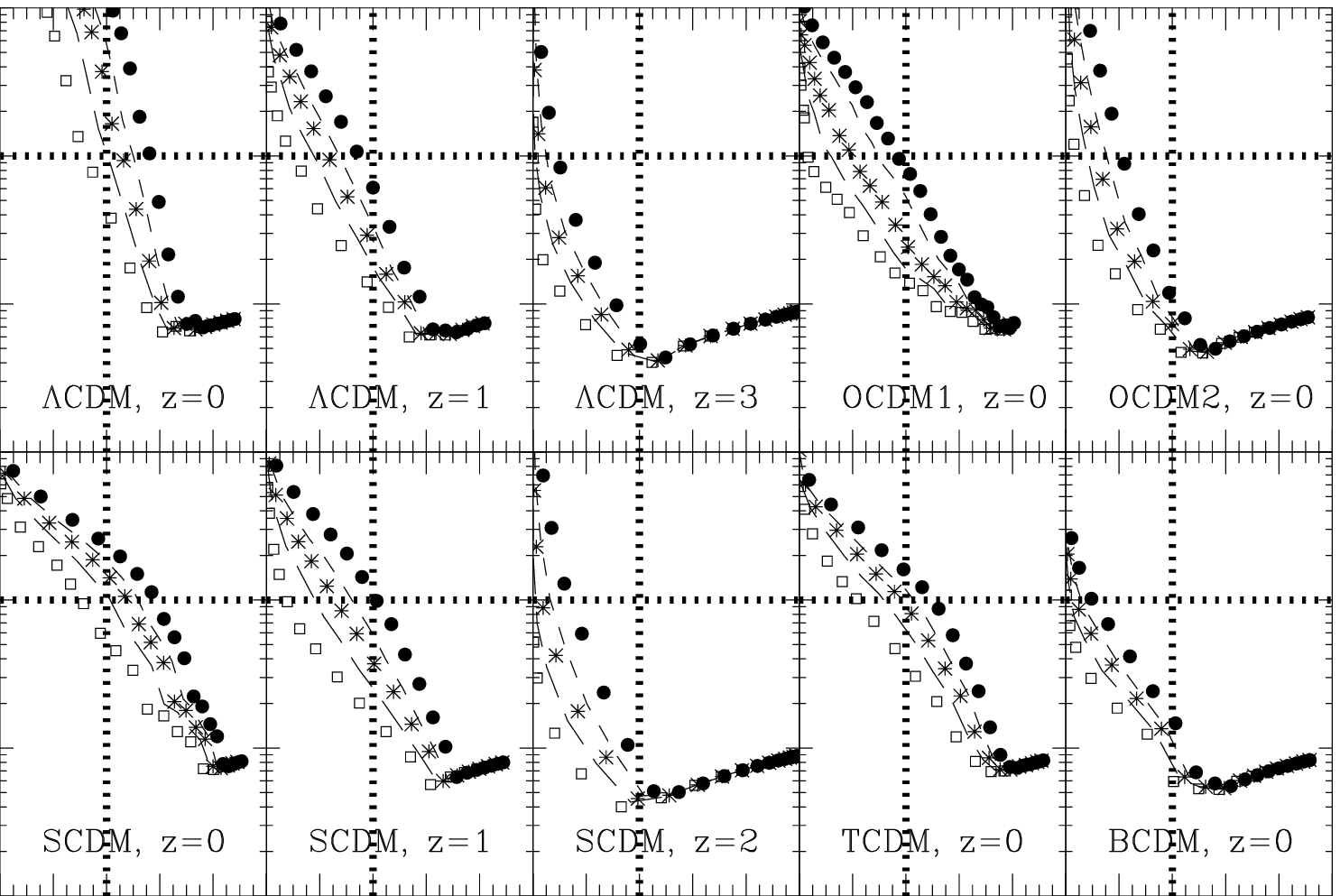,width=0.8\hsize}
}}
\vspace{0.8cm}
\caption[]{Overdensity $\delta_{ODR}$ in redshift space vs.
matter fraction $f_{ODR}$ for thresholds: $N_{th}=100$ (dots), 
$N_{th}=200$ (dashed line), $N_{th}=300$ (stars),
$N_{th}=500$ (long-dashed line), $N_{th}=1000$ (open squares).
Dotted lines show the observed parameters from LCRS.}
\end{figure}

In Fig.  3 we show the dependence of the mean overdensity of clusters on the
fraction of matter in the supercluster structures.  The different symbols belong
to different mass thresholds (here given in terms of the minimum number of
particles per cluster).  The dotted lines show the approximate characteristics
of the LCRS (Doroshkevich, 1996).  Obviously we get rich structures in the high
density regions only if the model is well evolved, in particular, OCDM2 and BCDM
cannot reproduce the degree of superclustering in the LCRS.

\begin{figure}
\centerline{\vbox{
\psfig{figure=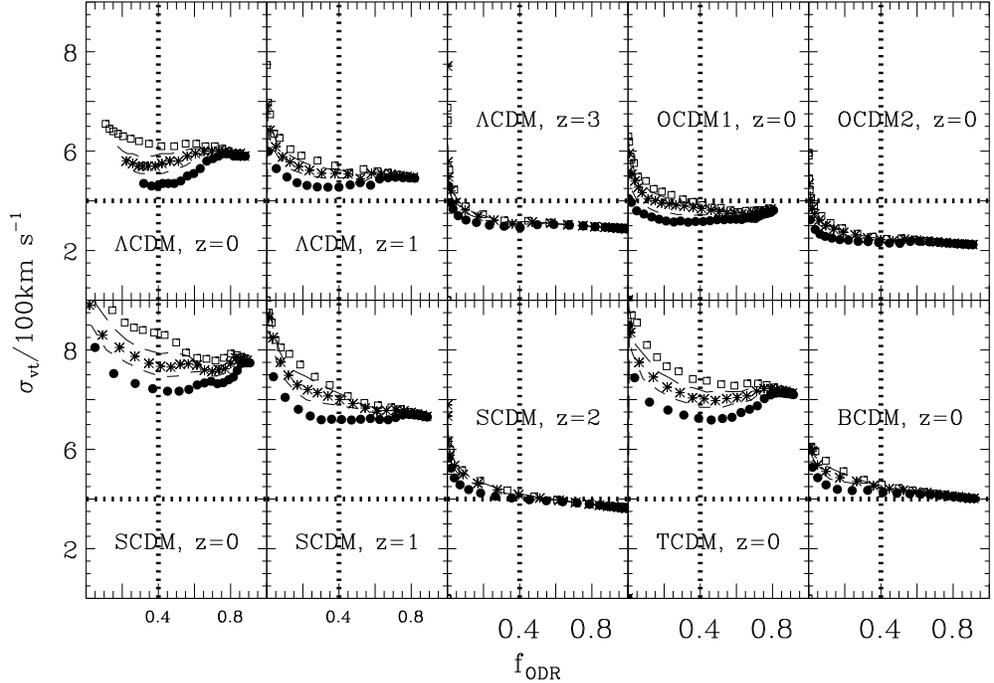,width=0.9\hsize}
}}
\vspace{0.8cm}
\caption[]{Velocity dispersion $\sigma_{vt}$ vs. matter fraction
concentrated within rich structure elements $F_{RSE}$ for the same 
richness thresholds as in Fig. 3. Dotted lines show the observed 
parameters in the LCRS.}
\end{figure}

In Fig.  4 we show the dependence of the cluster velocity dispersion along the
minor axis in dependence on the fraction of matter within clusters.  The
velocity dispersion can be estimated from the data either using the extension of
the superclusters along their smallest axis assuming virial equilibrium of the
transverse galaxy motion.  In this spirit, already Oort (1983) gave an estimate
of the observed velocity dispersion $\sigma_v^{obs} \approx (350 - 400)$ km/s.
The transverse velocity dispersion of superclusters in SCDM and TCDM is much to
high, i.e.  these models do not lead to the observed narrowness in the
supercluster distribution.  Also, $\Lambda$CDM shows a slight excess in the
velocity dispersion along the smallest axis.  The stablest superclustering is
produced in OCDM1.  It is remarkable that the OCDM2 models has a too low
velocity dispersion and a too low concentration of matter in superclusters.
This model is simply underevolved.

For SCDM and $\Lambda$CDM we show a few evolution steps.  It is obvious that the
supercluster structures are built up quite recently in both models.  Also the
transverse velocity dispersion develops strongly due to the nonlinear
gravitational clustering and the disintegration of the wall-like structures in
huge filaments and high-density clumps with high and isotropic velocity
dispersion.  The strong evolution of the superclustering up to redshift $z = 3$
has to be compared with some observational evidence of large matter
inhomogeneities at redshifts $z \approx 2 - 3$ (Connolly et al 1996, Cristiani
et al 1996, Williger et al.  1996).  Compared to our simulation results, such
structures must be very rare at high redshifts, or there must be a strongly
evolving bias of the galaxy formation up to these redshifts.

We use the eigenvalues of the tensor of inertia to estimate the degree of
anisotropic gravitational collapse in forming the supercluster structures.  The
mean length of the rich structure elements amounts $L_1 \approx 16,~~17,
~~\mbox{and}~~24 h^{-1}$Mpc for SCDM, OCDM1 and $\Lambda$CDM, respectively.
Along the other axes we find a collapse by a factor of two and $(4 - 5)$,
respectively.  This is consistent with the mean overdensity of 10 for the RSE,
and we expect an initial extension of the superclusters in the simulations of
$L_{init}^{sim} \approx (20 - 25) h^{-1}$Mpc (comoving coordinates), i.e.  no
collapse of the structures along the longest axis.  This estimate enters in the
following theoretical estimates.

\section{Theoretical estimates of supercluster parameters}

The reproduction of the observed characteristics of the RSE in simulations with
a standard CDM-like power spectrum verifies that the observed structure may have
formed during the nonlinear evolution of small initial perturbations, and that
its characteristics can be derived from the parameters of the initial power
spectrum of Gaussian fluctuations, and a specific cosmological model.
Demia\'nski ~\&~ Doroshkevich (1998) delivered an approximate theoretical
description based on Zel'dovichs nonlinear theory of gravitational instability.
The comparison of the initial size of structures, $L_{init}$, the wall
separations, $D_s$, and the mean velocity dispersions in rich supercluster
structures, $\sigma_u$, and the velocity dispersion along the smalles axis of
the superclusters, $\sigma_3$, are most interesting.  They are connected to the
length scale $l_0$ and evolution parameter $\tau_0$ of the power spectrum:
$$L_{init}^{th}\approx 2.9\tau_0 l_0,$$ 
and for the SCDM, OCDM1 and $\Lambda$CDM models we have
$$L_{init}^{th}\approx 23h^{-1}{\rm Mpc},~~\approx 9 h^{-1}{\rm
Mpc}, ~~\approx 16h^{-1}{\rm Mpc},$$ 
respectively.  For the OCDM1 and $\Lambda$CDM models $L_{init}^{th}\approx (0.6
- 0.7) L_{init}^{sim}$ is found.  The difference to our numerical results is
caused partly by the mass threshold used, and partly by the estimates of the
cluster boundary.  In the SCDM model we get $L_{init}^{th}\approx 1.5
L_{init}^{sim}$, it is strongly influenced by the small scale clustering.

A theoretical estimate to the mean wall separation for the mean rich
superclusters is given by
$$D_s^{th}\approx 6\cdot l_0\sqrt{\tau_0}, $$
and for the three models above we have
$$D_s^{th}\approx 56h^{-1}{\rm Mpc},~~\approx 49h^{-1}
{\rm Mpc},~~\approx 72h^{-1}{\rm Mpc},$$
respectively.  These values are consistent with the values of $D_s$ found with
the core-sampling method.

The theoretical model shows that for Gaussian perturbations the distribution of
the random velocity of structure elements is also Gaussian with a velocity
dispersion
$$\sigma_u^{th}\approx \sqrt{3}H_0(\beta-1)l_0\tau_0 =
(\beta-1)\sigma_{pvel} $$
where the parameter $\beta$ characterizes approximately the evolutionary stage
of perturbations, for SCDM $\beta=2$, otherwise 
$$\beta\approx (1+4\Omega_m)/(1+1.5\Omega_m)\approx 1.7~~
{\rm for ~~OCDM1}$$
$$\beta\approx (1+3.4\Omega_m)/(1+1.2\Omega_m)\approx 1.54~~
{\rm for} ~~\Lambda CDM. $$
For rich structure elements, $\sigma_u^{th}$ is lower by a factor $\approx
\sqrt{l_0/L_{init}^{th}} = (2.9\tau_0)^{-1/2}$.  Then for SCDM, OCDM1, and
$\Lambda$CDM, we have
$$\sigma_u\approx 950\,\mbox{km/s},\quad 385\,\mbox{km/s}, 
  \quad 490\,\mbox{km/s},$$
respectively.  These estimates are consistent to the simulations.

A theoretical estimate for the inner velocity dispersion in RSE along the
smallest axis $\sigma_3$ takes into account only the matter infall into RSE, and
it ignores the disruption of the pancakes into high-density clumps.  We have
$$\sigma_3^{th}\approx H_0(\beta-1){L_{init}\over 2\sqrt{3}} $$
and for a typical primordial size, $L_{init}^{th}$, given above, 
$$\sigma_3^{th}\approx 0.5(\beta-1)\sigma_{pvel}\approx 0.5\sigma_u. $$
This expression should be taken as a lower limit for the inner velocity
dispersion, and, really, for all models under consideration, we have
$\sigma_3^{th}\approx$ (0.5 -- 0.6) $\sigma_3$, i.e.  the influence of small
scale clustering on the properties of RSE is very important.

\section{Discussion}

In our analysis the properties of simulated spatial matter distributions were
studied for six cosmological models with CDM-like power spectra.  The
simulations were performed in large boxes in order to reproduce correctly the
mutual interaction of large and small scale perturbations, and to obtain a
representative sample of wall-like RSE.  The broad set of considered
cosmological models allows us to reveal the influence of main cosmological
parameters on the formation and evolution of the wall-like RSE, and to
discriminate between these models.  Our results show that the used methods are
effective, and they yield a description of the spatial matter distribution as a
whole and, in particular, the characteristics of the supercluster distribution.
The main results of our analysis can be summarized as following:

\begin{itemize}
\item{}The phenomenon of the strong matter concentration of galaxies in
wall-like rich structures can be reproduced with standard COBE normalized
CDM-like power spectra for suitable cosmological models and simulation
parameters.  An essential fraction of DM, $f_{rse} \approx$ 0.4, is compressed
nonlinearly on the scales $\sim$ (15 -- 20) $h^{-1}$Mpc that is less then the
mean separation of these RSE by a factor of $\sim$ 2 -- 3.

\item{} The comparison of observed and simulated parameters of the wall-like
supercluster distribution allows us to discriminate between different
cosmological models.  Only the low density models, the $\Lambda$CDM model with
$\Omega_m h \approx 0.15 - 0.25$, or the OCDM1 model with $\Omega_m h \approx$
0.25 -- 0.35 can reproduce the observed concentration of galaxies in rich
superclusters.  Probably, similar results can be obtained for a model with
a mixture of cold and hot dark matter.  A large scale bias between the spatial
distribution of DM and galaxies can also improve the predictions of some models.

\item{}Our results underline theoretical expectations concerning the epoch of
the supercluster formation.  At $z=1$ the fraction of matter accumulated by
superclusters with the chosen richness and overdensity drops by a factor $\sim
2$, and at $z = 3$, it becomes very small.

\item{} The simulations reproduces both the wall-like supercluster distribution
and the filamentary component of the low density regions.  Each component
represents an essential fraction of matter in the universe, and it is equally
important for the description of the joint network structure in the spatial
matter distribution.

\end{itemize}

Taken together only the low density models, best OCDM1, but also $\Lambda$CDM,
can reproduce the large sheetlike matter distribution observed in recent
redshift surveys.

\end{document}